\documentclass[a4paper,11pt]{article}
\usepackage{pos}
\usepackage{lineno}
\usepackage{hyperref}
\usepackage{graphicx}
\usepackage{pgfplots}
\usepackage{subcaption}

\usepackage{xcolor}
\usepackage{listings}
\usepackage{inconsolata}
\definecolor{EEBlue}{cmyk}{0.50, 0.30, 0, 0}            
\definecolor{EEBlueLight}{cmyk}{0.14, 0.04, 0.03, 0}    
\definecolor{EEBlueDark}{cmyk}{0.73, 0.55, 0.35, 0.32}  
\definecolor{EEOrange}{cmyk}{0.02, 0.41, 0.94, 0}       
\definecolor{EETangerine}{RGB}{255, 119, 0}
\definecolor{EEGold}{RGB}{255, 209, 0}
\definecolor{EEAzure}{RGB}{0, 136, 255}
\definecolor{EEUltramarine}{RGB}{0, 46, 255}

\RequirePackage{listings}
\lstdefinestyle{EEStyle}{
    backgroundcolor=\color{white},
    commentstyle=\color{EEBlueDark},
    keywordstyle=\bfseries\color{EEUltramarine},
    numberstyle=\tiny\color{EEBlue},
    stringstyle=\color{EETangerine},
    basicstyle=\ttfamily\small,
    breakatwhitespace=false,
    breaklines=true,
    captionpos=b,
    keepspaces=true,
    numbers=left,
    firstnumber=1, 
    stepnumber=1,  
    numbersep=5pt,
    showspaces=false,
    showstringspaces=false,
    showtabs=false,
    tabsize=2,
    frame=lB,
    frameround=tttt,
    inputpath=src
}
\title{IPbus extension for the ALFRED framework}

\author*[a]{Krystian Roslon}
\author[a]{Maciej Czarnynoga}
\author[b]{Sebastian Koryciak}
\author[c]{Monika Kutyla}
\author[d]{Ignacy Mermer}
\author[e]{Jacek Tomasz Otwinowski}
\author[b]{Wiktor Pierozak}
\author[b]{Pawel Grzegorz Russek}
\author[f]{Wladyslaw Trzaska}
\author[b]{Franciszek Urbanski}

\affiliation[a]{Faculty of Physics, Warsaw University of Technology, \\ Koszykowa 75, 00-662 Warsaw, Poland}
\affiliation[b]{AGH University of Krakow, al. Mickiewicza 30, 30-059 Kraków, Poland}
\affiliation[c]{Organisation Européenne pour la Recherche Nucléaire (CERN), \\ F-01631 Prévessin Cedex, France CH-1211 Genève 23, Geneva, Switzerland}
\affiliation[d]{Faculty of Mechatronics, Warsaw University of Technology, \\ sw. Andrzeja Boboli 8, 02-525 Warsaw, Poland}
\affiliation[e]{The Henryk Niewodniczański Institute of Nuclear Physics (IFJ) Polish Academy of Sciences (PAN) \\ ul. Radzikowskiego 152, 31-342 Kraków, Krakow, Poland}
\affiliation[f]{Department of Physics University of Jyvaskyla \\
P.O. Box 35, FIN-40351 Jyvaskyla, Finland}

\emailAdd{krystian.roslon@pw.edu.pl}

\date{August 2024}

\begin{document}

\abstract{We present a development of the ALFRED framework that includes support for the IPbus protocol, created for the Fast Interaction Trigger (FIT) detector in the ALICE experiment at CERN. This modification resolves the incompatibility between the current GBT-based slow-control protocols and the FIT electronics, which is based on the IPbus. A compatibility layer, named ALF IPbus, was created in C++ and integrated with DIM services, allowing for the transparent conversion of SWT frames into IPbus transactions. Benchmark tests indicated median execution times of less than 4.2 $\mu$s per frame for sequences above 512 words, comfortably meeting the ALICE Detector Control System requirements. Long-term stability tests conducted in the FV0 detector achieved an uptime close to 100\% from the moment of implementation on the FIT-FV0 detector. This solution enables FIT operation without firmware changes, ensures smooth integration with SCADA systems, and provides a scalable pathway for future migration to GBT-based control. The approach is general and can be applied to other ALICE subdetectors and high-energy physics experiments requiring protocol bridging. \\\\

\textbf{Keywords:} IPbus, FRED, ALFRED, protocol bridging, DCS}

\maketitle

\section{Introduction}
The Detector Control System (DCS) is a crucial component of every large-scale experiment, including ALICE\cite{ALICE:2008ngc} at the CERN Large Hadron Collider (LHC)\cite{Evans:2008zzb}. These complex experimental setups are not frozen on the commissioning day but evolve with developments in detector technologies and luminosity upgrades of the collider. Consequently, DCS must adapt accordingly. This paper describes our solution to the incompatibility problem between the ALFRED framework currently used by ALICE and the IPbus protocol employed by the front-end electronics of the new Fast Interaction Trigger detector (FIT)\cite{TRZASKA2020162116}. While the presented problem is specific to ALICE, the described solution is general and can be applied to other large-scale experiments that require protocol bridging between the existing GBT-based Single Word Transfer (SWT) communication and the IPbus, without firmware changes.\cite{Roslon:2025tym}
\section{ALICE Experiment}

\begin{figure}[ht]
    \centering
    \includegraphics[width=0.85\textwidth]{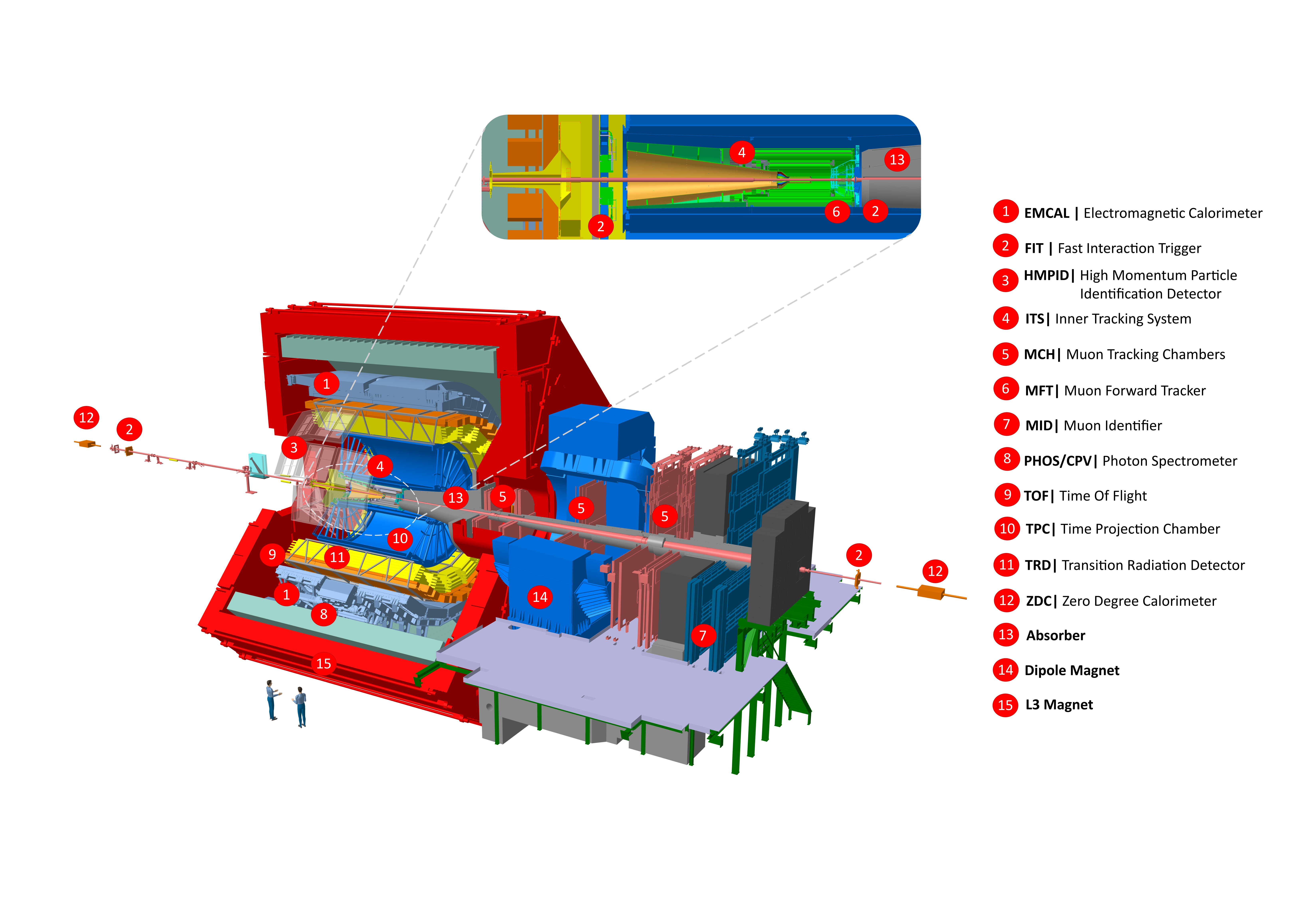}
    \caption{The ALICE experiment's schema.\cite{alice-detector}}
    \label{fig:Schema the ALICE experiment.}
\end{figure}

The ALICE (A Large Ion Collider Experiment) experiment at CERN is a highly specialised detector designed to investigate the behaviour of quark-gluon plasma (QGP) in ultra-relativistic heavy-ion collisions provided by the Large Hadron Collider (LHC)\cite{Niedziela:2017agp}. QGP represents a high--energy phase of strongly interacting matter, where quarks and gluons, ordinarily confined within hadrons such as protons and neutrons, are deconfined. Investigating QGP under such extreme conditions—characterised by temperatures that surpass 100,000 times those found at the core of the Sun—addresses fundamental questions related to Quantum Chromodynamics (QCD), the aspect of the Standard Model that describes strong interactions. A schematic view of the ALICE experiment is shown in Figure \ref{fig:Schema the ALICE experiment.}.

In the ALICE experiment, heavy nuclei, such as lead (208Pb), collide at velocities approaching the speed of light, generating a high-energy environment that mimics the state of the early universe just microseconds after the Big Bang. By analysing the formation, evolution, and eventual transition of QGP back into conventional hadronic matter, ALICE sheds light on confinement mechanisms and the collective dynamics of strongly interacting particles.

The ALICE setup measures 16 metres × 16 metres × 26 metres and weighs over 10,000  metric tons\cite{ALICE:2008ngc}. It comprises a set of specialised subsystems that utilise a variety of detection technologies, including silicon pixel detectors, scintillators, gas-based detectors, and Cherenkov radiation detectors, to monitor parameters such as particle momentum, energy, and trajectory across diverse spectra. The extracted parameters provide particle identification (PID) and momentum resolution, which are necessary to describe the dynamics and structure of QGP.

To accommodate the extraordinary collision rate increase during the high-intensity Run 3 and 4 of the LHC, ALICE underwent significant upgrades. These include new detector systems, such as the new Inner Tracking System (ITS), the Fast Interaction Trigger (FIT) \cite{Antonioli:2013ppp}, and high-speed electronics to process data rates over 1 TB/s \cite{Buncic:2015ari}, transmitted through thousands of optical links to an advanced computing farm for real-time processing. This data pipeline, a major technological feat, allows ALICE to handle particle yields orders of magnitude higher than in its original configuration. The detector can now read, analyse, and acquire data at speeds orders of magnitude above its original design parameters\cite{Buncic:2015ari}.

In addition to lead-lead collisions, ALICE records proton-proton, proton-nucleus, oxygen-oxygen, neon-neon and xenon-xenon collisions. In the present configuration, based on the L3 magnet and TPC tracking, ALICE is expected to operate through LHC Runs 3 and 4. Beyond that, a significant upgrade is needed. The proposed ALICE 3\cite{ALICE:2022wwr}, a next-generation multidetector, will have a superconducting magnet and silicon-based particle tracking.
\section{FIT detector setup}
\graphicspath{ {./images/} }

\begin{figure}[ht]
    \centering
    \includegraphics[width=0.85\textwidth]{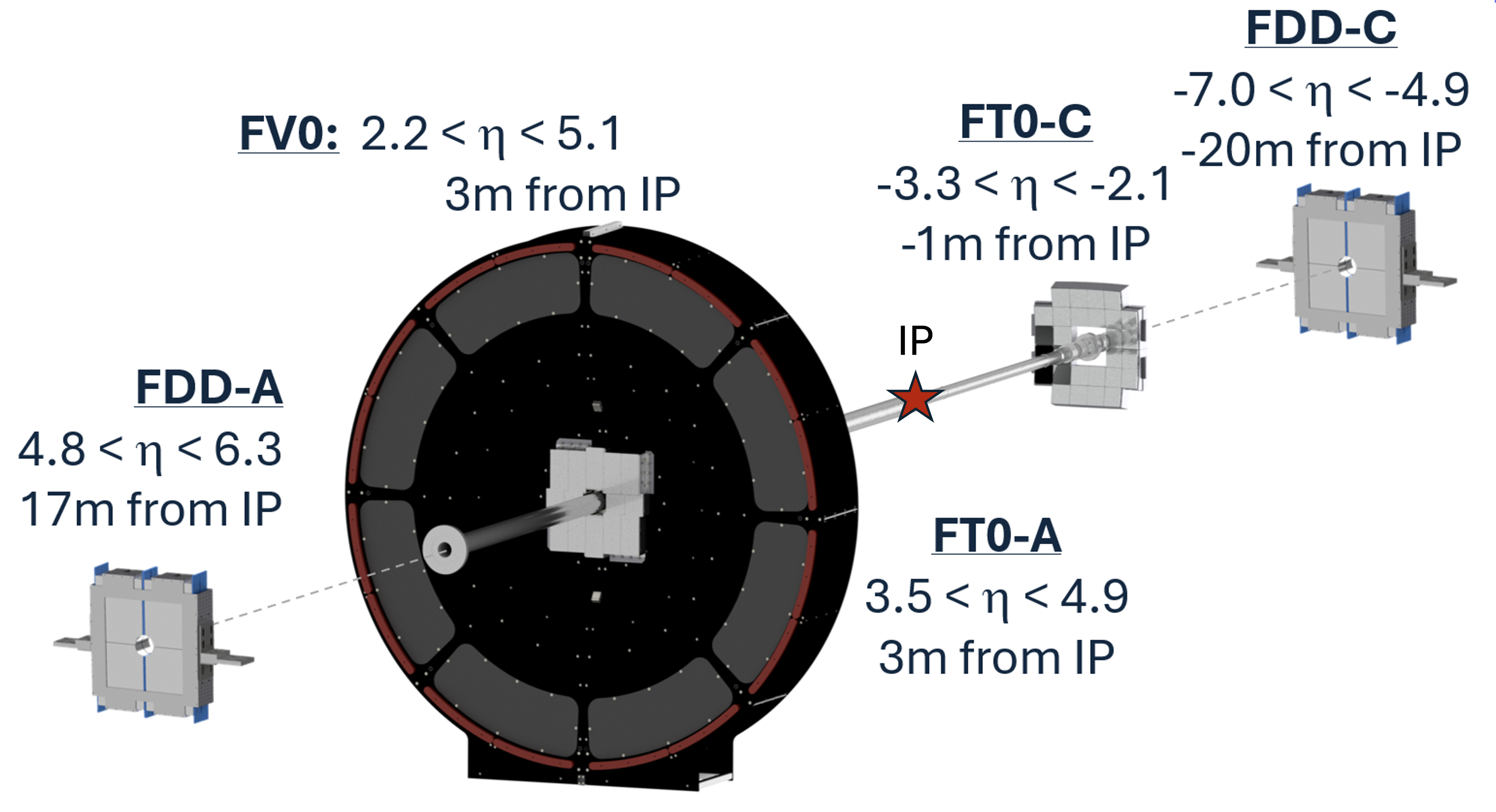}
    \caption{Schematic illustration of the FIT detector elements with the pseudorapidity coverage and distance from the interaction point listed.}
    \label{fig:FIT_overview}
\end{figure}

The ALICE Fast Interaction Trigger (FIT) consists of three detectors: FT0, FV0, and FDD, which were integrated into the ALICE experiment during the LHC Long Shutdown 2 (LHC LS2 - 2018-2022). As outlined in the Letter of Intent \cite{Abelev}, these detectors fulfill several 
functionalities \cite{Molander:20242y, SLUPECKI2022167021,Rytkonen:2021ZJ}, including:
\begin{itemize}
    \item Fast minimum bias collision trigger with latency < 425 ns;
    \item Precise collision time for Particle Identification with the Time-of-Flight detector;
    \item Forward multiplicity measurement;
    \item Diffraction physics measurements;
    \item Luminosity and background monitoring;
    \item Centrality and Event Plane determination;
    \item Veto trigger for UltraPeripheral Collision.
\end{itemize}
The FIT detectors are positioned in the forward regions of the ALICE on both sides of the Interaction Point (IP). The schematic arrangement of the detectors, along with their pseudorapidity coverage, is shown in Figure \ref{fig:FIT_overview}.

The FT0 detector consists of two Cherenkov arrays, one on each side of the Interaction Point (IP). The arrays include 52 modified Planacon XP85012 MCP-PMTs\cite{Grigoryev:2016ekg} coupled to 2 cm thick quartz radiators. Each radiator and the corresponding Planacon’s anodes are divided into four equal square sectors, yielding a total of 208 independent pixels, sized 7 $\text{cm}^2$. Cherenkov technology was chosen to deliver excellent time resolution (4.4 ps in Pb-Pb and 17 ps in pp collisions) and provide direction sensitivity required to suppress particle background originating from sources other than the IP\cite{TRZASKA2017463}.

The FV0 is a scintillator disc with an outer diameter of 148 cm, subdivided into five rings of eight 45-degree sections. Ring radii were calculated to provide uniform pseudorapidity coverage across the detection area. In the outermost ring, the neighbouring sections are read by the same photo-sensor, resulting in a total of 48 independent readout channels.\cite{TRZASKA2020162116} The utilisation of the light collection system without wavelength shifters provides an improved time resolution and reduced latency compared to the standard design\cite{grabski2024newfiberreadoutdesign}.

The FDD comprises scintillator arrays positioned on opposite sides of the IP, at a considerable distance from it. Each scintillator array consists of two layers of plastic scintillators, with each layer subdivided into four quadrants. Within each quadrant, two wavelength-shifting (WLS) bars are connected to individual photomultipliers through a bundle of clear optical fibres. The pseudorapidity coverage of the FDD allows for efficient tagging of diffractive and ultraperipheral events.\cite{RojasTorres:2020wS}

Regardless of the detector technology employed or the number of channels utilised, a uniform readout electronics system was implemented across all detectors, with the sole modification being the Analogue-to-Digital Converter (ADC) for scintillation detectors. The FIT's readout electronics consists of two types of modules: the Processing Module (PM) and the Trigger and Clock Module (TCM). The analogue signals generated in the detectors are processed within the PMs, where event timing is assessed, and input charge is integrated and subsequently digitised.\cite{FINOGEEV2020161920} Each PM accommodates up to 12 independent input channels. All PMs corresponding to a particular detector are connected to a single TCM, facilitating the exchange of "pre-trigger" and slow-control data, as well as the distribution of the LHC clock signals. In the TCM the trigger signals are formed in less than 225 ns \cite{Finogeev_2020}.

\begin{figure}[ht]
    \centering
    \includegraphics[width=0.85\textwidth]{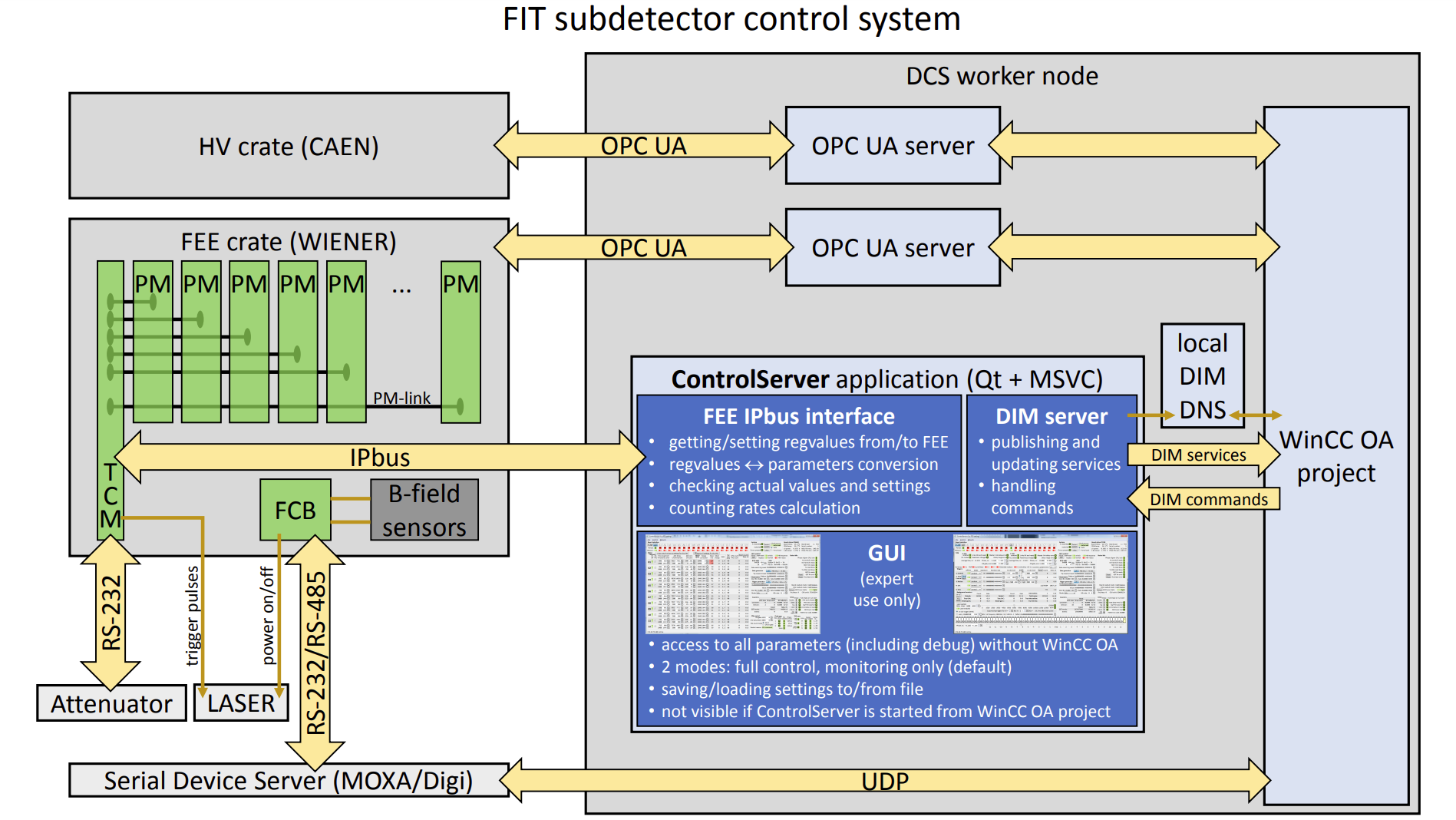}
    \caption{Schematic architecture of the FIT sub-detector control system.\cite{Sukhanov:2025fwu}}
    \label{fig:FIT_control}
\end{figure}

The overall architecture of the sub-detector control system is illustrated in Figure \ref{fig:FIT_control}. The top layer of the control system is based on the object-oriented Supervisory Control And Data Acquisition (SCADA) system SIMATIC WinCC Open Architecture (WinCC OA)\cite{WinCCOA}. WinCC OA interfaces with the CAEN high-voltage (HV) crate, which supplies FIT detectors, and the WIENER crate, housing the TCM and PMs, through an OPC Unified Architecture (OPC UA) server. Communication between WinCC OA and TCM takes place via a custom Control Server (CS) application. For communication between CS and WinCC OA, a local Distributed Information Management (DIM) server is used, through which DIM services and DIM commands are exchanged.\cite{Gaspar:2001fbw} The CS encompasses various functionalities, some of which are listed in Figure \ref{fig:FIT_control}, and can work as an independent application for the readout electronics control and monitoring. Data transmission between the TCM and the CS occurs over a 1 Gb Ethernet link utilising an IPbus, which is based on the UDP protocol.

\section{ALFRED framework}
\graphicspath{ {./images/} }

In the ALICE experiment at CERN, the integrated ALFRED framework\cite{FRED_TK}, consisting of the ALF (ALICE Low Level Frontend) and FRED (Frontend Device) systems, plays a crucial role in ensuring robust control and monitoring of custom detector electronics. Designed to operate on separate computers and network structures, ALF and FRED work in tandem to provide both a high-level and a low-level interface for managing ALICE subdetectors operations, supporting a streamlined and adaptable infrastructure for one of the world’s most complex experimental environments.

The ALF system acts primarily as an interface broker for the underlying electronics, granting remote access to detector data and controls via the DIM network protocol. ALF’s core function is to interface with the CRU (Common Readout Unit) cards\cite{Bourrion_2021}, which directly communicate with the detector modules via GBT (Gigabit Transceiver) links\cite{Bourrion_2021}. It accomplishes this by translating commands received from FRED over DIM into specific actions and ensuring atomic access to control sequences for efficient, conflict-free data handling. This ensures data integrity by completing command sequences in entirety before the next command is accepted, even if there are multiple remote requests. ALF’s simplicity and lightweight design allow it to manage the flow of information without performing complex data operations, reserving processing power for real-time communication.

The FRED system complements ALF by providing a high-level interface responsible for translating raw data from the detector into meaningful physical quantities. In addition, FRED allows dynamic configuration, enabling easy adaptation to different experimental setups. It maintains a flexible, scalable structure capable of supporting multiple ALF servers and CRU cards, which allows FRED to manage a large quantity of electronics simultaneously, making it ideal for distributed control systems like those in ALICE. Through this arrangement, WinCC OA, the SCADA system responsible for overseeing the entire ALICE detector network, can operate efficiently without needing detailed knowledge of the lower-level electronics, focusing instead on higher-level control and monitoring tasks.

\begin{figure}[ht]
    \centering
    \includegraphics[width=4.5 cm]{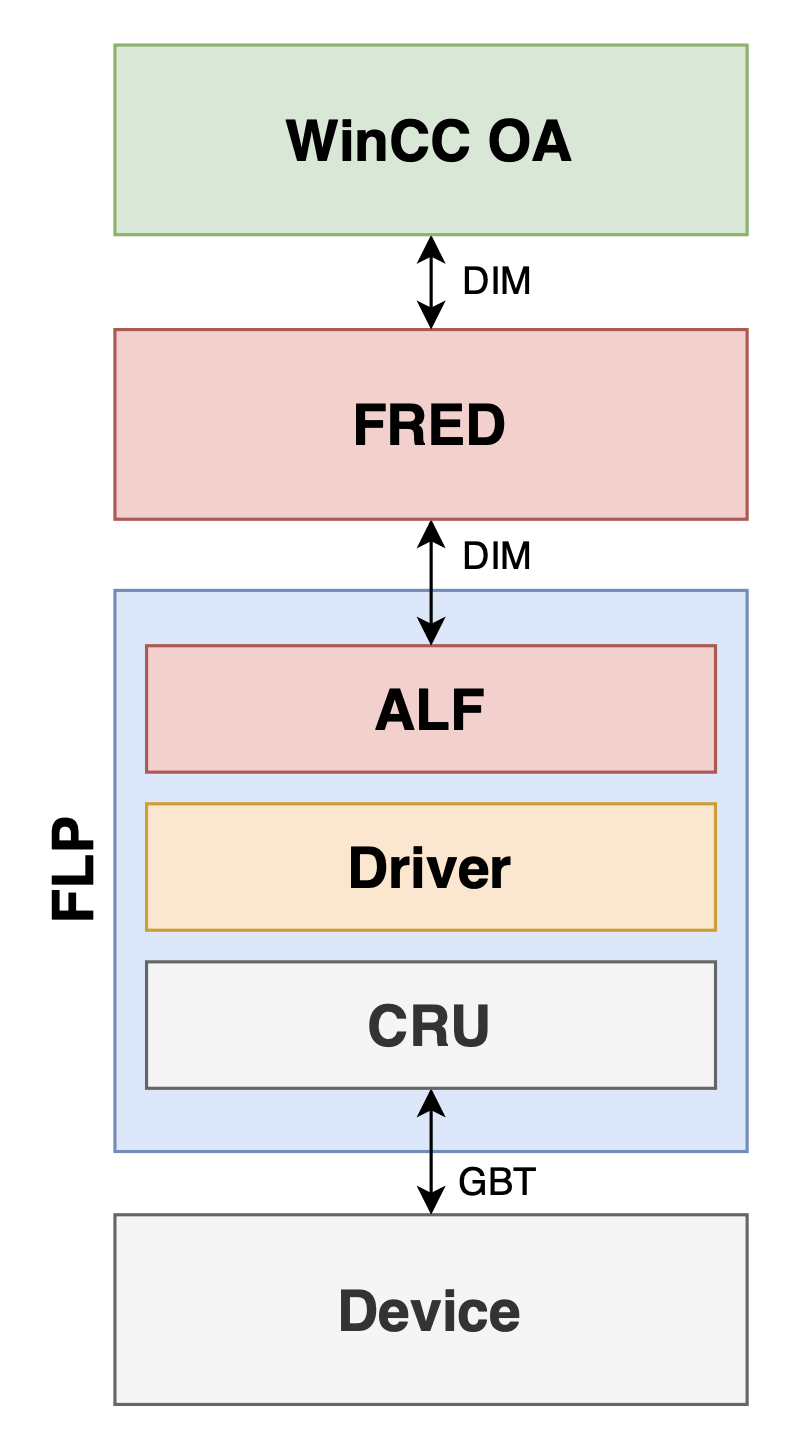}
    \caption{Flowchart showing the flow of data from the detector electronics, through ALF and FRED, up to the SCADA system.\cite{FRED_TK}}
    \label{fig:FRED}
\end{figure}

A key feature of ALFRED’s communication strategy is its reliance on the GBT protocol, providing a robust communication link between detector electronics and the control system. The GBT protocol is specifically designed for high-speed, bidirectional data transfer, achieving data rates of up to 4.8 Gb/s, which is crucial for supporting the high-throughput demands of the ALICE experiment.

ALFRED’s architecture also facilitates flexibility in protocol selection, allowing for a range of communication protocols within the detector setup. The framework can be adapted to work with protocols beyond GBT and DIM, such as CAN\cite{FRED_TK}, making it applicable to various environments that require high scalability and rapid data handling. Furthermore, the ALFRED framework’s unique capacity to separate the electronics interface from control logic is a notable advantage, allowing seamless integration of different detector components and protocols without requiring extensive reconfiguration of the SCADA system.
\section{IPbus protocol}

In the ALICE FIT detector setup, communication between the Frontend Electronics (FEE) and SCADA system is achieved through the IPbus protocol\cite{Finogeev_2020}. Communication protocols for data transfer in large-scale physics experiments, such as ALICE, must meet scalability and reliability requirements, enabling stable operation under various environmental conditions, while also providing flexibility for modification and expansion. A critical aspect of these systems is the provision of high-speed connections that can transfer large data packets in a single transaction. In this setup, a high-performance connection is established between the FEE and the network switch, supporting the coordination of multiple control computers within the detector infrastructure.

\begin{figure}[ht]
    \centering
    \includegraphics[width=13.5 cm]{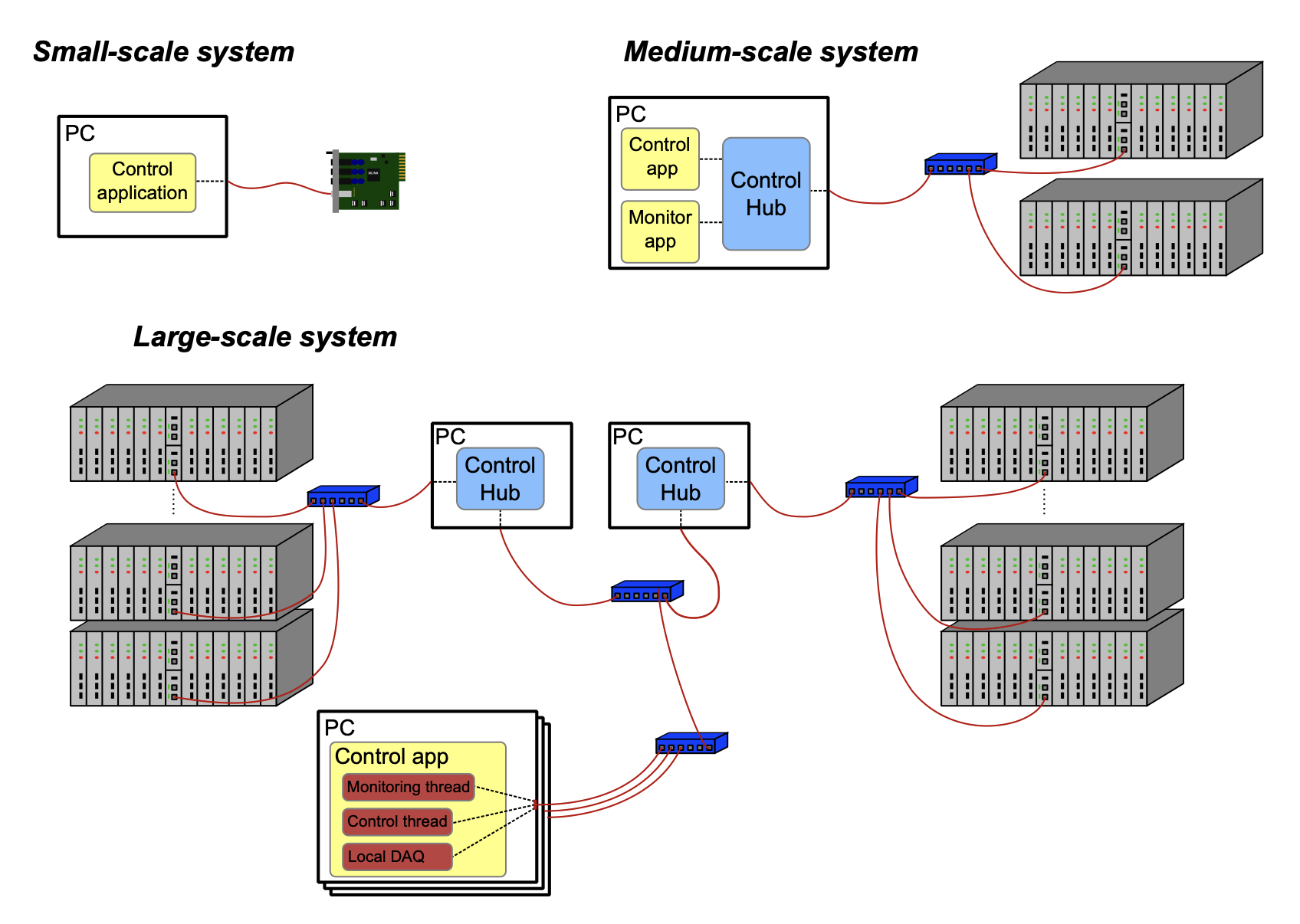}
    \caption{Example topologies of IPbus control systems, from small to large scale\cite{Larrea_2015}.}
    \label{fig:IPbus_topologies}
\end{figure}

Developed in 2009, the IPbus protocol simplifies and standardises data transfer in acquisition systems, particularly in settings that demand reliable, high-speed communication via TCP/IP protocols over Ethernet networks.\cite{Larrea_2015} Its primary strengths lie in its simplicity and efficiency, facilitating seamless integration with pre-existing network infrastructure without necessitating additional, costly hardware. By leveraging standard networking technologies, IPbus is adaptable across diverse environments, requiring no special modifications (See Fig. \ref{fig:IPbus_topologies}).

Software designed with IPbus support enables rapid development and immediate deployment of communication systems. This protocol has seen widespread adoption in high-energy physics projects, notably at CERN, where efficient management of the vast data volumes generated by particle detectors is essential. In the ALICE FIT detector, IPbus supports communication within the "Control Server" application, which oversees detector management and handles interactions between WinCC OA and the FEE, as well as within the "Histograms Reader" application, an expert tool for analysing detector data and identifying interference patterns. Additionally, IPbus underpins the ALF IPbus and ALFRED implementations, facilitating data transfer between the FEE and WinCC OA.\cite{ipbus_protocol_v2_0}

The IPbus architecture operates on a master-slave model, where the control software functions as the master, while the FEE acts as the slave. Communication is performed by sending IPbus packets through a transport protocol (recommended -- UDP).\cite{ipbus_protocol_v2_0}

The request and reply of an IPbus packet always begin with a 32‐bit (single IPbus word) header. Its structure is displayed in Table \ref{tab:ipbus-header}. The packet ID is a sequential number, used for the IPbus reliability mechanism. The byte order qualifier allows to easily distinguish big and little endian packets. The packet type value for standard packets containing transactions (control packets) is 0x0.

\begin{table}[h]
    \centering
    \begin{tabular}{|c|c|c|c|c|}
    \hline
    31-28 & 28-24 & 23-8 & 7-4 & 3-0 \\
    \hline
    Protocol version  & Rsvd.  & Packet ID  & Byte-order qualifier  & Packet type  \\
    \hline
    0x2 & 0 & 0x0 - 0xffff & 0xf & 0x0 - 0x2 \\
    \hline
    \end{tabular}
    \caption{IPbus packet header format}
    \label{tab:ipbus-header}
\end{table}

After the header, each IPbus packet can contain several transactions. The protocol supports several transaction types:
\begin{itemize}
    \item READ: incrementing and non-incrementing;
    \item WRITE: incrementing and non-incrementing;
    \item RMWbits (Read-Modify-Write bits): enables selective setting or clearing of bits within a register;
    \item RMWsum (Read-Modify-Write sum): allows for additive operations on register values.
\end{itemize}
The incrementing transaction variants perform operations on consecutive addresses when multi-word. The non-incrementing variants perform the operation repeatedly on a single address. Each transaction begins with a transaction header, outlined in Table \ref{tab:ipbus-transaction-header}.

\begin{table}[h]
    \centering
    \begin{tabular}{|c|c|c|c|c|}
    \hline
    31-28 & 27-16 & 15-8 & 7-4 & 3-0 \\
    \hline
    Protocol version  & Transaction ID  & Words  & Type ID  & Info code  \\
    \hline
    0x2 & 0 & 0x0 - 0xffff & 0xf & 0x0 - 0x2 \\
    \hline
    \end{tabular}
    \caption{IPbus transaction header format}
    \label{tab:ipbus-transaction-header}
\end{table}

In Tables \ref{tab:read-request}-\ref{tab:rmwsum-response}, formats of all IPbus transaction types are shown.

\begin{table}[h]
\centering
\begin{tabular}{|c|c|c|c|c|c|}
\hline
Word 0 & Version = 2 & Transaction ID & Words = READ\_SIZE & Type ID & InfoCode = 0xf \\
\hline
Word 1 & \multicolumn{5}{|c|}{BASE\_ADDRESS} \\
\hline
\end{tabular}
\caption{Incrementing read (type ID 0x0) and non-incrementing read (type ID 0x2) request format}
\label{tab:read-request}
\end{table}

\begin{table}[h]
\centering
\begin{tabular}{|c|c|c|c|c|c|}
\hline
Word 0 & Version = 2 & Transaction ID & Words = READ\_SIZE & Type ID = 0 & InfoCode = 0 \\
\hline
Word 1 & \multicolumn{5}{|c|}{Data read from BASE\_ADDRESS} \\
\hline
Word 2 & \multicolumn{5}{|c|}{Data read from BASE\_ADDRESS + 1} \\
\hline
\ldots & \multicolumn{5}{|c|}{\ldots} \\
\hline
Word n & \multicolumn{5}{|c|}{Data read from BASE\_ADDRESS + (READ\_SIZE - 1)} \\
\hline
\end{tabular}
\caption{Incrementing read (type ID 0x0) and non-incrementing read (type ID 0x2) response format}
\label{tab:read-response}
\end{table}

\begin{table}[h]
\centering
\begin{tabular}{|c|c|c|c|c|c|}
\hline
Word 0 & Version = 2 & Transaction ID & Words = WRITE\_SIZE & Type ID & InfoCode = 0xf \\
\hline
Word 1 & \multicolumn{5}{|c|}{BASE\_ADDRESS} \\
\hline
Word 2 & \multicolumn{5}{|c|}{Data for BASE\_ADDRESS } \\
\hline
Word 3 & \multicolumn{5}{|c|}{Data for BASE\_ADDRESS + 1 } \\
\hline
\ldots & \multicolumn{5}{|c|}{\ldots} \\
\hline
Word n & \multicolumn{5}{|c|}{Data for BASE\_ADDRESS + (WRITE\_SIZE - 1) } \\
\hline
\end{tabular}
\caption{Incrementing write (type ID 0x1) and non-incrementing read (type ID 0x3) request format}
\label{tab:write-request}
\end{table}

\begin{table}[h]
\centering
\begin{tabular}{|c|c|c|c|c|c|}
\hline
Word 0 & Version = 2 & Transaction ID & Words = WRITE\_SIZE & Type ID & InfoCode = 0 \\
\hline
\end{tabular}
\caption{Incrementing write (type ID 0x1) and non-incrementing read (type ID 0x3) response format}
\label{tab:write-response}
\end{table}

\begin{table}[h]
\centering
\begin{tabular}{|c|c|c|c|c|c|}
\hline
Word 0 & Version = 2 & Transaction ID & Words = 1 & Type ID & InfoCode = 0xf \\
\hline
Word 1 & \multicolumn{5}{|c|}{BASE\_ADDRESS} \\
\hline
Word 2 & \multicolumn{5}{|c|}{AND term} \\
\hline
Word 3 & \multicolumn{5}{|c|}{OR term} \\
\hline
\end{tabular}
\caption{RMWbits (type ID 0x4) request format}
\label{tab:rmwbits-request}
\end{table}

\begin{table}[h]
\centering
\begin{tabular}{|c|c|c|c|c|c|}
\hline
Word 0 & Version = 2 & Transaction ID & Words = 1 & Type ID & InfoCode = 0 \\
\hline
Word 1 & \multicolumn{5}{|c|}{Content of BASE\_ADDRESS as read before the modify/write} \\
\hline
\end{tabular}
\caption{RMWbits (type ID 0x4) response format}
\label{tab:rmwbits-response}
\end{table}

\begin{table}[h]
\centering
\begin{tabular}{|c|c|c|c|c|c|}
\hline
Word 0 & Version = 2 & Transaction ID & Words = 1 & Type ID & InfoCode = 0xf \\
\hline
Word 1 & \multicolumn{5}{|c|}{BASE\_ADDRESS} \\
\hline
Word 2 & \multicolumn{5}{|c|}{ADDEND} \\
\hline
\end{tabular}
\caption{RMWsum (type ID 0x5) request format}
\label{tab:rmwsum-request}
\end{table}

\begin{table}[h]
\centering
\begin{tabular}{|c|c|c|c|c|c|}
\hline
Word 0 & Version = 2 & Transaction ID & Words = 1 & Type ID & InfoCode = 0 \\
\hline
Word 1 & \multicolumn{5}{|c|}{Content of BASE\_ADDRESS as read before the summation} \\
\hline
\end{tabular}
\caption{RMWsum (type ID 0x5) response format}
\label{tab:rmwsum-response}
\end{table}

To establish communication between the application and electronics, an IPbus host device, located on the TCM board within the FIT electronics, is configured with an IP address and port number. Client application requests are directed to this address, and responses are returned to the client’s source address. IPbus also allows for the consolidation of multiple requests into a single packet, effectively minimising transmission delays.

\begin{figure}[ht]
    \centering
    \includegraphics[width=13.5 cm]{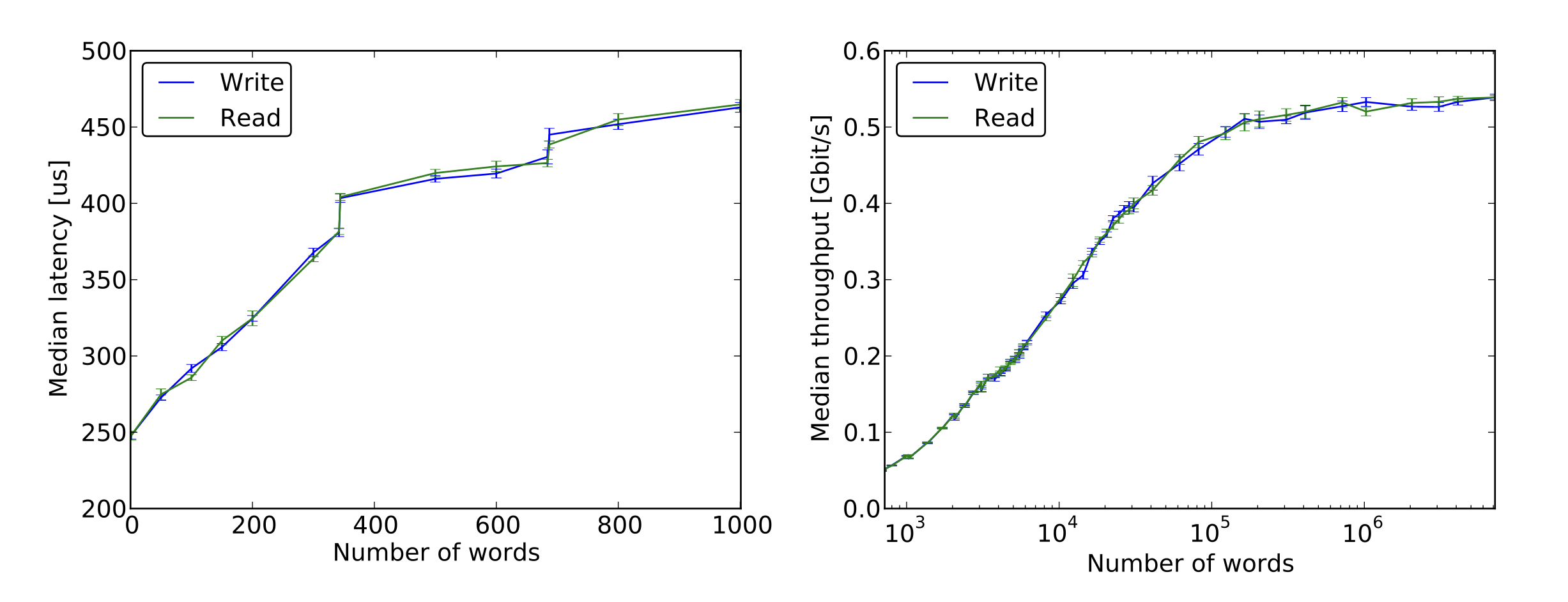}
    \caption{The median write/read latency and throughput as a function of depth, for one software client controlling one IPbus device\cite{Larrea_2015}.}
    \label{fig:IPbus_time}
\end{figure}

The performance of the IPbus protocol can be evaluated by measuring transmission delays relative to the number of words in read and write operations. For instance, packets containing fewer than 10 words typically experience a delay of approximately 250 $\mu$s. The following graph (See Fig. \ref{fig:IPbus_time}) illustrates these delays as a function of word count in both read and write operations, demonstrating IPbus’s effectiveness in handling varying data loads.

\section{IPbus extension to ALFRED}

\subsection{Genesis}
The ALFRED framework assumes a GBT-based slow-control communication protocol is used to control the FEE. The GBT standard defines two such protocols: SWT (Single-Word Transaction) and SCA (Slow Control Adapter) \cite{Gabrielli:2009xli} \cite{Bourrion_2021}. The GBT-SCA requires a dedicated ASIC and has a strict message format, whereas SWT does not define the format but leaves 76 bit for user-defined messages.

The change in communication protocol created an additional difficulty in the system upgrade, as certain procedures rely on specific operation types (e.g. an atomic read-modify-write). Changing this logic would require a profound change in the firmware, which was not possible within the upgrade schedule. It was therefore decided that an intermediate compatibility layer would be needed to facilitate the upgrade.

For the message format, SWT communication was chosen, as it is more flexible than SCA. The FIT SWT protocol was defined.

\subsection{FIT SWT}
The GBT standard allows for the transmission of 80-bit slow control frames. Each SWT frame has a special 4-bit identifier in the first bits, and the remaining bits are reserved for a user-defined message. 
FIT SWT utilises the last 68 bits as visualised in Table \ref{tab:fit-swt}. Transactions are based on 4-byte words, which resemble IPbus communication.

Besides standard read and write transactions, it was necessary to implement two types of atomic read-modify-write operations for IPbus compatibility. The first type, called RMWbits, applies an AND mask followed by an OR mask to the register data. The second type, RMWsum, sums the operation argument with register data. RMWbits could not be implemented as a single frame and had to be split into two SWT frames.

To accommodate all requirements, a multiword read operation was defined, apart from the single-word read operation. There are two variants of such operation: incrementing (multiple registers) and non-incrementing (FIFO). Both are limited to 1024 words per transaction to align with hardware limits on SWT FIFO on the CRU (for future shift to ALF).
Table \ref{tab:fit_swt_transactions} summarises all transaction types.

\begin{table}[h]
    \centering
    \begin{tabular}{|c|c|c|c|c|c|}
        \hline
        SWT header (4b) & Not used (8b) & Transaction type (4b) & Address (32b) & Data (32b) \\
        \hline
    \end{tabular}
    \caption{FIT SWT frame}
    \label{tab:fit-swt}
\end{table}

\begin{table}[h]
\centering
\begin{tabular}{|c|c|c|}
        \hline
         Frame type & Code & Data field meaning   \\
         \hline
         Read & 0x0 & - \\
         Write & 0x1 & Data to write \\
         RMW AND & 0x2 & AND mask in RMWbits operation \\
         RMW OR & 0x3 & OR mask in RMWbits operation \\
         RMW Sum & 0x4 & Value to sum with register data \\
         Incrementing block read & 0x8 & Number of words to read \\
         Non-incrementing block read & 0x9 & Number of words to read \\
         \hline
         \end{tabular}
    \caption{FIT SWT transaction types summary}
    \label{tab:fit_swt_transactions}
\end{table}

\subsection{ALF IPbus}
As mentioned earlier, the new DCS solution requires a compatibility layer to work with the current hardware and firmware setup. This was achieved by implementing a piece of software that exposes an interface identical to that of ALF to FRED, while communicating with FEE using IPbus. This enabled a switch of DCS systems without firmware intervention. Furthermore, the compatibility with ALF ensures that the future data path change to GBT will be transparent to higher-level software. ALF IPbus name was chosen to emphasise its role in connecting the standard ALF interface with the IPbus protocol. ALF IPbus is written in C++, and utilises the DIM interface used in the ALICE DCS system. Communication via IPbus is performed through a custom IPbus C++ implementation. 

ALF IPbus takes as input an SWT frame sequence written in CRU-aware format and translates it to IPbus packets. The packets are then sent to the FEE, and the IPbus response is translated back to a FRED-compatible format.

One of the biggest challenges was to properly simulate CRU operations corresponding to sending SWT frames (by writing them to a dedicated register) and reading the SWT FIFO that contains responses from the FEE. The granularity of GBT transactions (one frame per packet) differs from that of IPbus (multiple transactions per packet). This incompatibility is hidden for FRED -- SWT transactions are translated into IPbus transactions and packaged efficiently while remaining unnoticeable in the ALF IPbus response.


\subsection{Testing}
To ensure the correct operation of ALF IPbus, a flexible testing framework\cite{IPbus_ALF_tester} has been developed, consisting of two tools: Mock and Generator. Mock simulates an IPbus slave device. Its address space can be configured via a .csv file, and it supports all IPbus operations, as well as response randomisation (for simulating errors). The generator generates FIT SWT sequences and tracks the test progress. Mock can track Generator communication to allow for automated testing and validation with simulated errors. Both components can be configured via a single TOML file, which defines the operations to be performed during the tests.

An example TOML configuration file for the tester is shown in List. \ref{lst:toml}.

\begin{lstlisting}[language=C,
    backgroundcolor=\color{EEGold!5!white},
    caption={An example TOML configuration file for the tester},
    label={lst:toml}]
[global]
name = "Example config"
register_file = "registers/fee.csv"
rng_seed = 36
alf = { name = "ALF", serial = 0, link = 0 }

[[tests]]
name = "Write test"
enabled = true
registers = [
    { begin = 0x0, end = 0xf }, # Register block, begin to end inclusively
    0x1004, 
    0x1005, 
]
operations = [
    { type = "read" },
    { type = "write", data = [0xdeadbeef] },
    { type = "rmw_bits", data = [0xfffff000, 0x00000001] },
]
randomise_response = true
randomise_operations = true
split_seq = true
repeats = 100
wait = 10000

[[tests]]
name = "Load test"
enabled = true
registers = [0x1004]
operations = [
    { type = "rmw_sum" },
    { type = "read" },
]
repeats = 1024
wait = 0
randomise_operations = true
\end{lstlisting}

The \texttt{global} section defines the general settings for the test, including the file from which the register map is read, the seed for the pseudo-random number generator used for response randomisation, and the parameters of the ALF IPbus link being tested.

The \texttt{tests} table contains specifications for tests, which are executed in order. Tests define \texttt{registers}, on each of which all the specified \texttt{operations} are performed (single sequence for all registers, or separate for each one, as determined by \texttt{split\_seq}). The tests can be repeated number of times, waiting microseconds between each operation. Randomisation of the order of operations within one register can also be performed. Moreover, by enabling the \texttt{randomise\_response} flag, Mock can simulate failure of the operation execution, and Generator can check whether the error handling is correct. This tracking of test execution is enabled by the fact that both tools use the same type of pseudo-random number generator, seeded with the specified value, so their decisions as to whether a given operation should fail or not are identical.

This arrangement makes it possible to define many different test procedures in a simple way, testing numerous edge cases and performance-critical areas, and has proved to be very useful in ensuring the smooth and stable operation of ALF IPbus.

\section{Benchmarks}

\begin{figure}[h!]
\centering 

\begin{subfigure}{0.48\textwidth}
    \centering
\begin{tikzpicture}
    \begin{axis}[
        title={Total execution time},
        xlabel={SWT frames},
        ylabel={Total RET [µs]},
        xmode=log,
        ymode=linear,
        grid=major,
        width = \textwidth,
        legend pos=north west
    ]
    \addplot+[
        color=green,
        mark=*,
        error bars/.cd,
            y dir=both,
            y explicit,
    ]
    coordinates {
        (1.0, 328.01) +- (0, 47.955062) (2.0, 344.6) +- (0, 64.647) (4.0, 330.24) +- (0, 55.414) (8.0, 329.15) +- (0, 60.66) (16.0, 374.55) +- (0, 72.45) (32.0, 328.96) +- (0, 50.19) (64.0, 398.26) +- (0, 70.37) (128.0, 432.7) +- (0, 76.24) (256.0, 813.17) +- (0, 101.4) (512.0, 1359.71) +- (0, 169.96) (1024.0, 2636.76) +- (0, 246.54)
    };
    \addlegendentry{Read}

    \addplot+[
        color=blue,
        mark=square*,
        error bars/.cd,
            y dir=both,
            y explicit,
    ]
    coordinates {
        (1, 352.81) +- (0, 66.857) (2, 331.42) +- (0, 63.47) (4, 350.77) +- (0, 68.29) (8, 340.49) +- (0, 65.72) (16, 332.82) +- (0, 55.05) (32, 341.35) +- (0, 59.29) (64, 390.89) +- (0, 72.08) (128, 651.98) +- (0, 86.58) (256, 861.33) +- (0, 83.55) (512, 1505.09) +- (0, 171.1287) (1024, 2742.41)
    };
    \addlegendentry{Write}

    \addplot+[
        color=red,
        mark=triangle*,
        error bars/.cd,
            y dir=both,
            y explicit,
    ]
    coordinates {
        (1, 344.15) +- (0, 39.577) (2, 327.31) +- (0, 49.52) (4, 334.45) +- (0, 63.24) (8, 338.29) +- (0, 55.48) (16, 350.52) +- (0, 64.98) (32, 370.56) +- (0, 64.4) (64, 406.45) +- (0, 76.21) (128, 780.79) +- (0, 108.3) (256, 1188.77) +- (0, 196.14705) (512, 2123.34) +- (0, 184.5182) (1024, 4327.18) +- (0,539.5993)
    };
    \addlegendentry{RMW Bits}

    \end{axis}
\end{tikzpicture}

    \caption{Total processing time of SWT sequence}
    \label{fig:operation-characteristic-total-time}
\end{subfigure}
\hfill 
\begin{subfigure}{0.48\textwidth}
    \centering
\begin{tikzpicture}
    \begin{axis}[
        title={Execution Time Per Frame},
        xlabel={SWT frames},
        ylabel={Execution Time [µs]},
        xmode=log,
        ymode=log,
        grid=major,
        legend pos=north east,
        width=\textwidth, 
    ]
    \addplot+[
        color=red,
        mark=*,
        error bars/.cd,
            y dir=both,
            y explicit,
    ]
    coordinates {
        (1.0, 328.01) +- (0, 47.955062) (2.0, 172.3) +- (0, 32.32348) (4.0, 82.56) +- (0, 13.853568000000001) (8.0, 41.14375) +- (0, 7.582793124999999) (16.0, 23.409375) +- (0, 4.5273731249999996) (32.0, 10.28) +- (0, 1.5687280000000001) (64.0, 6.2228125) +- (0, 1.09957096875) (128.0, 3.38046875) +- (0, 0.59563859375) (256.0, 3.176445313) +- (0, 0.39610273046875) (512.0, 2.655683594) +- (0, 0.33196044921875) (1024.0, 2.574960938) +- (0, 0.24075884765625002)
    };
    \addlegendentry{Read}
    \addplot+[
        color=blue,
        mark=square*,
        error bars/.cd,
            y dir=both,
            y explicit,
    ]
    coordinates {
        (1, 352.81) +- (0, 66.857495) (2, 165.71) +- (0, 31.733465000000002) (4, 87.6925) +- (0, 17.073729750000002) (8, 42.56125) +- (0, 8.1632) (16, 20.80125) +- (0, 3.4405267499999996) (32, 10.6671875) +- (0, 1.85289046875) (64, 6.10765625) +- (0, 1.1262518125) (128, 5.09359375) +- (0, 0.67642925) (256, 3.3645) +- (0, 0.3263) (512, 2.9396) +- (0, 0.3342)
    };
    \addlegendentry{Write}
    \addplot+[
        color=green,
        mark=triangle*,
        error bars/.cd,
            y dir=both,
            y explicit,
    ]
    coordinates {
        (1, 344.15) +- (0, 39.57725) (2, 163.655) +- (0, 24.7610015) (4, 83.6125) +- (0, 15.811123749999998) (8, 42.28625) +- (0, 6.934945000000001) (16, 21.9075) +- (0, 4.0616505) (32, 11.58) +- (0, 2.012604) (64, 6.35078125) +- (0, 1.190771484375) (128, 6.099921875) +- (0, 0.8460591640624999) (256, 3.364570313) +- (0, 0.32636332031250004) (512, 4.14714) +- (0, 0.3603) (1024, 4.2257) +- (0, 0.5269)
    };
    \addlegendentry{RMW Bits}

    \end{axis}
\end{tikzpicture}
    \caption{Total execution time per single SWT frame}
    \label{fig:operation-characteristic-time-per-frame}
\end{subfigure}

\caption{Characteristics of ALF IPbus performance for multiple size of SWT sequence}
\label{fig:operation-characteristic}
\end{figure}
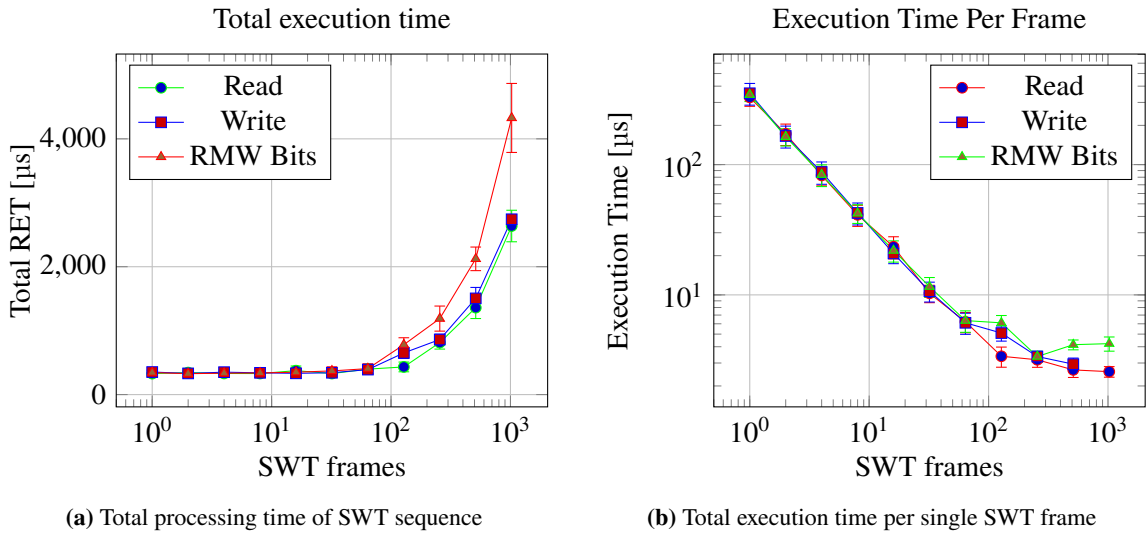

From the point of view of system stability, it was crucial to know how ALF IPbus performs across a wide range of SWT sequence sizes and for different operation types. The programme was benchmarked against sequences from 1 to 1024 frames of three types: read, write, and read–modify–write (RMW) bits. Figure \ref{fig:operation-characteristic-total-time} presents the total execution time. The total processing time remains almost constant for sequences of up to 64 operations and increases from 128 frames onward. The operation cost for RMW bits grows more rapidly than for read and write operations, as a single RMW-bit request requires two SWT frames and thus about twice the processing effort.

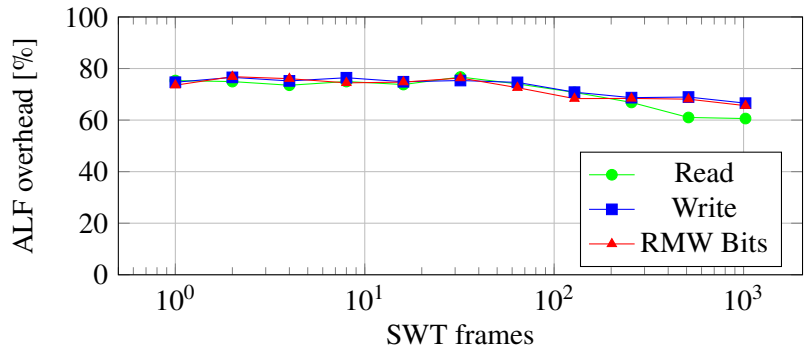
\begin{figure}[h!]
    \centering
    \begin{tikzpicture}
\begin{axis}[
    xlabel={SWT frames},
    ylabel={ALF overhead [\%]},
    ymin=0,
    ymax=100,
    xmode=log, 
    grid=major,
    width=0.7\textwidth,
    height= 5cm,
    legend pos=south east,
    yticklabel style={
        /pgf/number format/fixed,
        /pgf/number format/precision=2
    },
    scaled y ticks=false
]
\addplot[
    color=green,
    mark=*,
]
coordinates {
(1,75.3361178) (2,74.93325595) (4,73.51320252) (8,74.95366854) (16,73.78988119) (32,76.70537451) (64,74.03957214) (128,70.75803097) (256,66.84703076) (512,61.01595193) (1024,60.59557942)
};
\addlegendentry{Read}
\addplot[
    color=blue,
    mark=square*,
]
coordinates {
(1, 74.68042289) (2, 76.56749744) (4, 75.194571940000005) (8, 76.42221504) (16, 74.86929872000001) (32, 75.344953860000005) (64, 74.627132950000005) (128, 70.83039357) (256, 68.70421324999999) (512, 68.9354125) (1024, 66.56869573)
};
\addlegendentry{Write}
\addplot[
    color=red,
    mark=triangle*,
]
coordinates {
(1, 73.52898445) (2, 76.83847117) (4, 76.06219166) (8, 74.439090719999995) (16, 74.728974100000005) (32, 76.322322970000005) (64, 72.57473244000001) (128, 68.33335468) (256, 68.40768189) (512, 68.113443910000005) (1024, 65.59884266)
};
\addlegendentry{RMW Bits}
\end{axis}
\end{tikzpicture}
\caption{Translation overhead as percent of total processing cost}
\label{fig:alf-translation-overhead}
\end{figure}

As the IPbus packet size is limited to a single Ethernet MTU, increasing the SWT sequence size adds, beyond the translation itself, the burden of multiple IPbus packet exchanges. As seen in Figure \ref{fig:alf-translation-overhead}, starting at 256 frames (where the size of a single IPbus packet is first exceeded), the share of translation overhead—which includes protocol translation and updating the DIM service—slightly decreases.

Figure \ref{fig:operation-characteristic-time-per-frame} shows a clear decreasing trend in per-frame execution time as sequence size increases, leveling off near the upper end of the range at approximately 2.5 µs for read requests and 4.2 µs for RMW operations.

\begin{figure}[h!]
\centering
\begin{subfigure}{0.48\textwidth}
    \centering
    \begin{tikzpicture}
        \begin{axis}[
            title={FIFO Read Performance},
            xlabel={SWT frames},
            ylabel={Time [µs]},
            xmode=linear,
            ymode=linear,
            grid=major,
            legend pos=north west,
            width=\textwidth,
        ]
        \addplot+[
            color=blue,
            mark=*,
            error bars/.cd,
                y dir=both,
                y explicit,
        ] coordinates {(512, 4799.4) +- (0, 94.06823999999999) (1023, 8291.09) +- (0, 2585.990971) (2046, 18969.36) +- (0, 339.551544) (3069, 27497.0) +- (0, 896.4021999999999) (4092, 37675.27) +- (0, 1005.929709) (5115, 47384.83) +- (0, 1170.405301) (6400, 58626.71) +- (0, 1612.234525)};
        \addlegendentry{Read}

        \addplot+[
            color=red,
            mark=square*,
            error bars/.cd,
                y dir=both,
                y explicit,
        ] coordinates {(512, 4166.4) +- (0, 75.82848) (1023, 7996.4) +- (0, 139.13735999999997) (2046, 15693.0) +- (0, 252.6573) (3069, 23569.4) +- (0, 381.82428) (4092, 32287.25) +- (0, 742.60675) (5115, 39251.45) +- (0, 714.37639) (6400, 49589.4) +- (0, 669.4569)};
        \addlegendentry{Block Read}
        \end{axis}
    \end{tikzpicture}
    \caption{Total processing time for FIFO readout.}
    \label{fig:fifo-readout-performance-time}
\end{subfigure}
\hfill 
\begin{subfigure}{0.48\textwidth}
    \centering
    \begin{tikzpicture}
        \begin{axis}[
            title={FIFO read - Translation overhead},
            xlabel={SWT frames},
            ylabel={ALF Overhead [\%]},
            ymin = 0,
            xmode=linear,
            grid=major,
            legend pos=south east,
            width=\textwidth,
        ]
        \addplot+[color=blue, mark=*] coordinates {(512, 21.74646831) (1023, 19.08072401) (2046, 18.248533419999998) (3069, 18.01440157) (4092, 18.264925509999998) (5115, 18.31862645) (6400, 17.59704749)};
        \addlegendentry{Read}

        \addplot+[color=red, mark=square*] coordinates {(512, 16.31384409) (1023, 12.917062679999999) (2046, 11.582234119999999) (3069, 11.27987984) (4092, 13.026566210000002) (5115, 11.21184058) (6400, 11.53875627)};
        \addlegendentry{Block Read}
        \end{axis}
    \end{tikzpicture}
    \caption{Translation overhead percentage for FIFO readout.}
    \label{fig:fifo-readout-performance-overhead}
\end{subfigure}
\caption{Performance and overhead analysis of FIFO readout methods.}
\label{fig:fifo-readout-performance}
\end{figure}
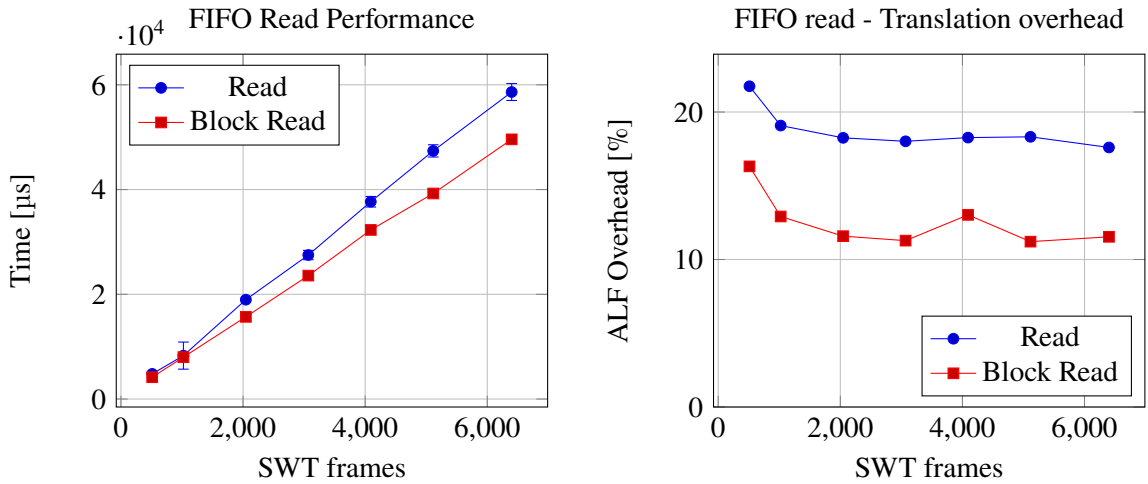

The most resource-intensive operation in FIT DCS is histogram reading. Histograms are retrieved either from a dedicated FIFO or, less frequently, from a range of registers. From the perspective of the front-end electronics, FIFO readout imposes a heavier load, as shown in Figure \ref{fig:fifo-readout-performance-overhead} - translation overhead accounts for only a small fraction of the total operation cost. It is opposite to non-FIFO case, where translation overweights IPbus transaction. Moreover, Figure \ref{fig:fifo-readout-performance-overhead} shows advantage of block read over single read - block read incur less overhead than single reads because the input sequence is shorter, reducing processing effort.

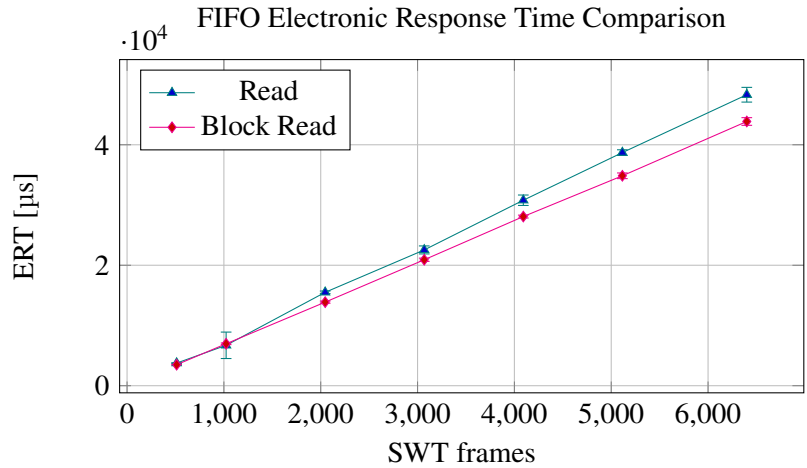
\begin{figure}[h!]
\centering
\begin{tikzpicture}
    \begin{axis}[
        title={FIFO Electronic Response Time Comparison},
        xlabel={SWT frames},
        ylabel={ERT [µs]},
        xmode=linear,
        ymode=linear,
        grid=major,
        legend pos=north west,
        height = 6cm,
        width=0.7\textwidth,
    ]
    \addplot+[
        color=teal, 
        mark=triangle*,
        error bars/.cd,
            y dir=both,
            y explicit,
    ] coordinates {(512, 3755.7) +- (0, 73.23615) (1023, 6709.09) +- (0, 2194.5433390000003) (2046, 15507.73) +- (0, 212.455901) (3069, 22543.58) +- (0, 665.03561) (4092, 30793.91) +- (0, 865.308871) (5115, 38704.58) +- (0, 464.45496) (6400, 48310.14) +- (0, 1222.2465419999999)};
    \addlegendentry{Read}
    \addplot+[
        color=magenta, 
        mark=diamond*,
        error bars/.cd,
            y dir=both,
            y explicit,
    ] coordinates {(512, 3486.7) +- (0, 56.484539999999996) (1023, 6963.5) +- (0, 138.57365000000001) (2046, 13875.4) +- (0, 180.38019999999997) (3069, 20910.8) +- (0, 267.65824) (4092, 28081.33) +- (0, 238.69130500000003) (5115, 34850.64) +- (0, 470.48364) (6400, 43867.4) +- (0, 636.0773)};
    \addlegendentry{Block Read}

    \end{axis}
\end{tikzpicture}
\caption{Electronic Response Time (ERT) of FIFO readout for both read methods}
\label{fig:electronic-response-time-fifo}
\end{figure}

Figure \ref{fig:fifo-readout-performance-time} shows a substantial performance difference between the read methods. This gap stems from both the higher processing cost of the input sequence and the longer IPbus readout for non-block reads. A difference in the electronic response time (ERT) is visible in Figure \ref{fig:electronic-response-time-fifo}. The SWT block read is translated into an IPbus block read in an optimal way—two half-packet-size requests—which improves performance on the FEE side.

\section{Implementation in FIT}
\begin{figure}[h!]
    \centering
    \includegraphics[width=1.0\textwidth]{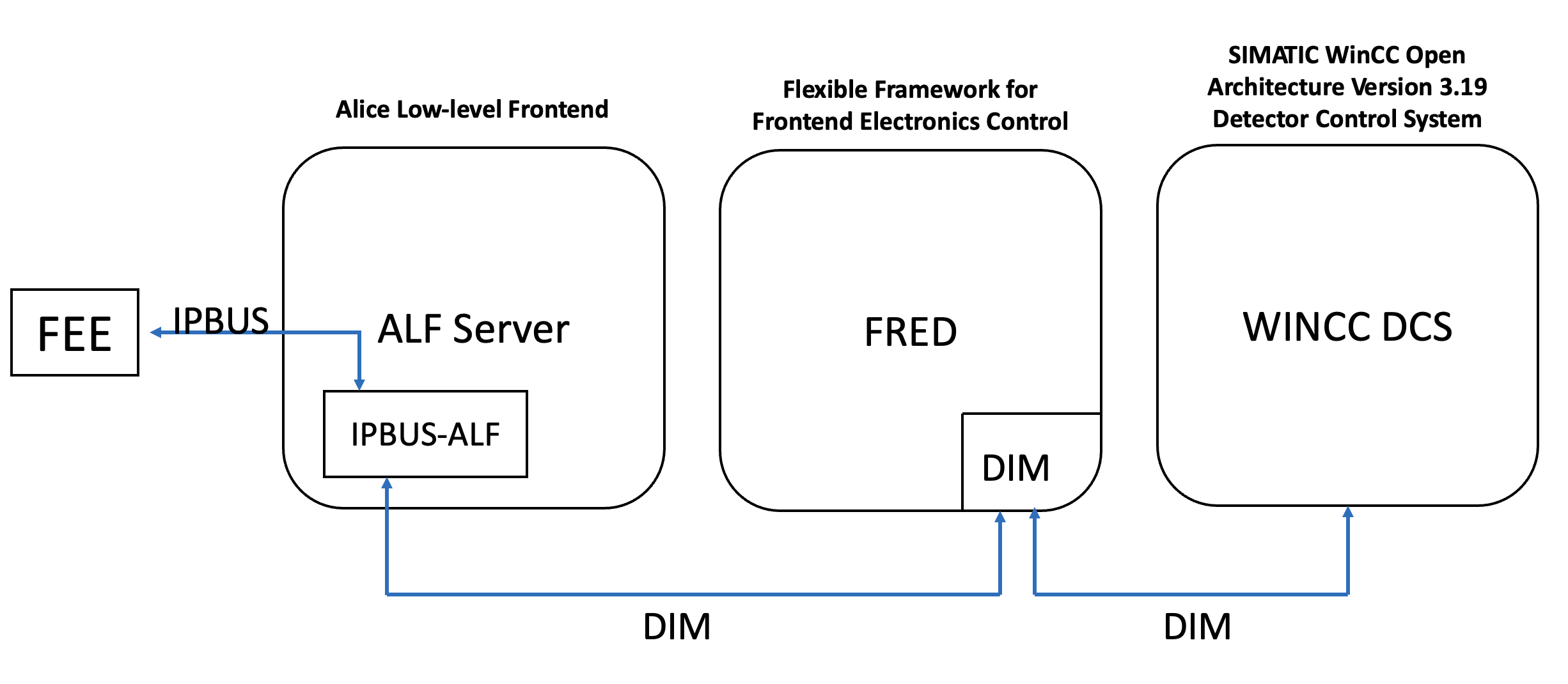}
    \caption{ALFRED implementation for FV0 schematic, with the IPbus extension \cite{Roslon:2025lxb}}
    \label{fig:schematDCS_CS}
\end{figure}

The solution described above has been implemented for the FV0 detector. ALF IPbus is running on a RHEL9 (Red Hat Enterprise Linux 9) node, which is connected via IPbus to the FEE. FRED, running on a separate RHEL9 node, communicates via DIM with ALF IPbus. Moreover, it communicates, also via DIM, with the WinCC project, which is running on the Worker Node. This arrangement is shown in Figure \ref{fig:schematDCS_CS}. 

So far, this solution has proved to be reliable and performant, without major issues affecting the long-term stability of the system. This arrangement allows for easy extension of the system, for instance, via implementing additional calibration procedures. The FV0 detector's implementation currently manages 48 independent readout channels through ALF IPbus. Continuous operation tests conducted from the moment of implementation on the detector demonstrated an uptime close to 100\%. The average DIM response time observed between FRED and ALF IPbus was 1670 $\mu s$, while the configuration procedure, consisting of read-modify-write operations, was completed in less than 1100 $\mu s$ per board. This confirms the system's stability under real experimental conditions and provides sufficient performance capacity for future expansion to FT0 and FDD, which is planned for the end of 2025.

\section{Conclusion}

The IPbus extension to ALFRED enabled FIT operation without requiring firmware modifications, while maintaining compatibility with upper-level SCADA systems. The system exhibited stable long-term performance since the moment of implementation on the FIT-FV0 detector.  This methodology is versatile and can be adapted for other ALICE subdetectors or experiments with analogous requirements. Future endeavours will focus on optimising the SWT→IPbus translation pipeline, aiming to decrease overhead, and completing the deployment to all FIT components by 2026, as well as preparing the hardware version of this device.

\section{Acknowledgments}
This work was supported by the Polish Ministry of Science and Higher Education under agreements no. 5452/CERN/2023/0, 2022/WK/01, 2023/WK/07 and ``The Excellence Initiative - Research University'' programme.

\bibliographystyle{unsrt}
\bibliography{bibliography}

\end{document}